\documentclass[preprint,nofootinbib,aps,superscriptaddress,eqsecnum]{revtex4-1} %\documentclass[11pt, a4paper]{article}
\pdfoutput=1
 \usepackage{graphicx}
 \usepackage{amsmath}
\usepackage{amsfonts}
\usepackage{amssymb}
\usepackage{hyperref}
\usepackage{gensymb}

\usepackage{caption}
\usepackage{subcaption}
\def\bea{\begin{eqnarray}}
\def\eea{\end{eqnarray}}
 \def\be{\begin{equation}}
\def\ee{\end{equation}}

\begin{document}

\title{ Implications of the Dark LMA solution and Fourth Sterile Neutrino for Neutrino-less Double Beta Decay }

 \author{K. N. Deepthi}
\email[Email Address: ]{nagadeepthi.kuchibhatla@mechyd.ac.in}
\affiliation{School of Natural Sciences, 
Mahindra Ecole Centrale, Hyderabad - 500043, India}

 \author{Srubabati Goswami}
\email[Email Address: ]{sruba@prl.res.in}
\affiliation{Theoretical Physics Division, 
Physical Research Laboratory, Ahmedabad - 380009, India}

 \author{Vishnudath K. N.}
\email[Email Address: ]{vishnudath@prl.res.in}
\affiliation{Theoretical Physics Division, 
Physical Research Laboratory, Ahmedabad - 380009, India}

  \author{Tanmay Kumar Poddar}
\email[Email Address: ]{tanmay@prl.res.in}
\affiliation{Theoretical Physics Division, 
Physical Research Laboratory, Ahmedabad - 380009, India}
\affiliation{Discipline of Physics, Indian Institute of Technology, Gandhinagar - 382355, India}

\begin{abstract}

We analyze the effect of the Dark-large mixing angle (DLMA) 
solution on the effective Majorana mass ($m_{\beta\beta}$) governing neutrino-less double beta decay ($0\nu\beta\beta$) in the presence of a sterile neutrino. 
We consider the 3+1 picture, comprising of one additional sterile neutrino. 
We have checked that the MSW resonance in the sun can take place in the DLMA parameter space in this scenario. Next we investigate how the values of the 
solar mixing angle $\theta_{12}$ corresponding to the DLMA region alter the 
predictions of $m_{\beta\beta}$ by including a sterile neutrino in the analysis.
We also compare our results with three generation cases for both standard large mixing angle (LMA) and DLMA. Additionally, we evaluate the discovery sensitivity of the future ${}^{136}Xe$ experiments in this context.

\end{abstract}
 
\maketitle

\section{Introduction}

The standard three flavour neutrino oscillation picture has been corroborated by the data from decades of experimentation on neutrinos. 
However some exceptions to this scenario have been reported over the years,  calling for the necessity of transcending beyond the three neutrino paradigm. 
The first among these signatures came from the LSND $\bar{\nu_\mu} \rightarrow \bar{\nu_e}$ oscillation data \cite{Aguilar:2001ty}, which could be explained by invoking additional neutrino states (sterile) that mix with active neutrinos \cite{Goswami:1995yq,Okada:1996kw,Bilenky:1996rw,
Bilenky:1999ny,Maltoni:2004ei}. This result was supported  by the hints obtained : from the appearance data of $\bar{\nu_\mu} \rightarrow \bar{\nu_e}$ and $\nu_\mu \rightarrow \nu_e$ at MiniBooNE experiment \cite{AguilarArevalo:2007it, AguilarArevalo:2008rc, AguilarArevalo:2010wv, Aguilar-Arevalo:2013pmq,Aguilar-Arevalo:2018gpe}, from the reactor neutrino anomaly \cite{Declais:1994su,Mention:2011rk} where a deficit in the 
$\bar{\nu_e}$ reactor flux has been reported by short baseline(SBL) oscillation data and also from the missing neutrino flux at 
GALLEX \cite{Anselmann:1994ar, Hampel:1997fc,Giunti:2010zu} and SAGE 
\cite{Abdurashitov:1996dp} source experiments. However, accelerator experiments like KARMEN \cite{Armbruster:2002mp}, ICARUS\cite{Antonello:2013gut} , NOMAD\cite{Astier:2003gs} have not found a positive signal. There are also disappearance experiments using reactors and accelerators as neutrino sources which have not reported any evidences of sterile neutrino \cite{Giunti:2019aiy}. The allowed region from the global analysis including all these data have been obtained in \cite{Gariazzo:2017fdh,Dentler:2018sju}. Several new experiments are planned to test the sterile neutrino hypothesis \cite{Conrad:2016sve}.

The basic question whether the neutrinos are  Dirac particles or lepton number violating Majorana particles (for which particles and antiparticles are the same) remains as a major puzzle in neutrino physics. Since oscillation experiments do not help us to determine the nature of the neutrinos, one has to rely on studying the processes in which total lepton number is violated. In this regard, neutrino-less double beta decay ($0\nu \beta\beta$) process $ (\,\,\, X_Z^A \,\,\rightarrow \,\, X_{Z+2}^A \,\, + \,\, 2e^- $) stands as a promising probe to establish the Majorana nature of neutrinos. $0\nu \beta\beta$ decay has not been observed so far and there are several ongoing and upcoming experiments that search for this signal. The best limit on the half life of $0\nu \beta\beta$ decay is $T_{1/2} > 1.07 \times 10^{26}$ years coming from the KamLAND-Zen experiment using ${}^{136}Xe$ \cite{KamLAND-Zen:2016pfg}. This gives a bound on the effective Majorana mass ($m_{\beta\beta}$) as,
$$ m_{\beta\beta} \,\,\leq \,\, 0.061 - 0.165 ~ \textrm{eV}.$$ The range corresponds to the uncertainty in nuclear matrix elements (NME).

This process is suppressed by the proportionality of the transition amplitude to the effective Majorana mass $m_{\beta\beta}$, which in turn depends on the lowest neutrino mass, neutrino mass ordering, mixing angles and Majorana phases. However, the predictions for $m_{\beta\beta}$ are known to change substantially in a 3+1 mixing scenario when an additional sterile neutrino is introduced \cite{Bilenky:2001xq,Benes:2005hn,Goswami:2005ng,Barry:2011wb,Li:2011ss,
Rodejohann:2012xd,Girardi:2013zra,
Giunti:2015kza,Jang:2018zug,Huang:2019qvq}. It is also well known that in the presence of non-standard interactions (NSI), solar neutrino data admits a new solution for $\theta_{12} > 45^{\circ}$, known as the dark large mixing angle (DLMA) solution \cite{Miranda:2004nb,Escrihuela:2009up,Farzan:2017xzy}. This is nearly a degenerate solution with $\Delta m^2_{21}\simeq 7.5\times 10^{-5}\textrm{eV}^2$ and $\sin^2\theta_{12}\simeq 0.7$. The DLMA parameter space was shown to be severely constrained from neutrino-nucleus scattering data from COHERENT experiment \cite{Esteban:2018ppq}. However the bound depends on the mass of the light mediator \cite{Denton:2018xmq}. In this context, the effect of the  DLMA solution on $0\nu\beta\beta$ for the standard three generation picture has been studied recently in ref.~\cite{N.:2019cot} where it was shown that  the prediction
for $m_{\beta\beta}$  
remains unchanged for the inverted mass scheme whereas for normal hierarchy, it 
becomes higher for the Dark-LMA parameter space and shifts to the 
``desert region'' between the two. This region can be tested in the
next generation experiments.

In this work, we have studied the implications of the Dark-LMA solution to the solar neutrino problem for $0\nu\beta\beta$ in the presence of one sterile neutrino as introduced to explain the LSND/MiniBooNE results (see references \cite{Giunti:2019aiy,Boser:2019rta} for recent reviews on the status of eV scale sterile neutrinos.). 
In this case, $m_{\beta\beta}$ depends on  the third mass-squared difference $\Delta m^2_{LSND}$, the mixing angle $\theta_{14}$  and an additional Majorana phase $\gamma/2$, in addition to the two mass squared differences $\Delta m_{21}^2$ and $\Delta m_{31}^2$, two mixing angles $\theta_{12}$ (degenerate LMA or DLMA solutions) and $\theta_{13}$ and the Majorana phases $\alpha/2$ and $\beta/2$.
Depending on the values of these parameters, there can be enhancement or cancellation of the $0\nu\beta\beta$ decay rate.

It has to be noted that the sum of masses of all the neutrino  species is highly constrained from cosmology, which does not allow an eV scale sterile neutrino (see \cite{Boser:2019rta} for a recent review on the status of light sterile neutrinos and the cosmological bounds). 
To avoid the cosmological constraints, one can invoke ``secret neutrino interactions" which can dynamically suppress the production of sterile neutrinos in the early universe by finite temperature effects \cite{Chu:2018gxk}. One may also avoid the cosmological constraints by assuming a very low reheating temperature $(\sim MeV)$ after inflation \cite{Gelmini:2004ah,Yaguna:2007wi,deSalas:2015glj}.

The rest of the paper is organized as follows. In the next section, we discuss the DLMA solution and the MSW resonance condition in the presence of one sterile neutrino. In section-\ref{section3}, we discuss the implications of the sterile neutrino and the DLMA solution for $0\nu\beta\beta$ process. The 
discovery sensitivity of $0\nu\beta\beta$ process in the new allowed parameter space is discussed in section-\ref{section4} 
in the context of ${}^{136}Xe$ based experiments. Finally, we summarize our results in section-\ref{section5}.

\section{DLMA solution in 3+1 neutrino framework}\label{section2}

In the 3+1 neutrino framework, the neutrino mixing matrix $U$ is a $4\times4$ unitary matrix which can be parametrized by three active neutrino mixing angles $\theta_{12}$, $\theta_{13}$, $\theta_{23}$, three active-sterile mixing angles $\theta_{14}$, $\theta_{24}$ and $\theta_{34}$ and the Dirac CP violating phases $\delta_{CP}, \delta_{14}, \delta_{24}$. Hence the $4\times 4$ unitary matrix is given by,
\begin{equation}
U=R_{34}\tilde{ R_{24}}\tilde{ R_{14}}R_{23}\tilde{ R_{13}}R_{12}P \label{eq:pmns},
\end{equation}
where $P=diag(1,e^{i\alpha/2},e^{i(\beta/2+\delta_{CP})}, e^{i(\gamma/2+\delta_{14})} )$, and $\alpha/2,~\beta/2,~\gamma/2$ are the Majorana phases.
The Dirac CP phases $\delta_{CP}$, $\delta_{14}$ and $\delta_{24}$ are
associated with $\tilde{ R_{13}}$, $\tilde{ R_{14}}$ and $\tilde{R_{24}}$
respectively. The Majorana phases can take values in the range $0-\pi$.
The rotation matrices $R$ and $\tilde{R}$ are given in the Eqn. (15) of reference \cite{Chakraborty:2019rjc}. The Majorana phase matrix comes into play while studying 
$0\nu \beta\beta$ process, but they are not relevant for oscillation studies. In Table \ref{oscdata}, we have given the $3\sigma $ ranges of the mixing angles and mass squared differences in the three generation \cite{Esteban:2018azc} as well as four generation schemes\cite{Gariazzo:2017fdh}. Similar analysis can also be found in references \cite{deSalas:2017kay,Capozzi:2017ipn} for three generation case and in \cite{Dentler:2018sju} for the four generation case.

%%%%%%%%%%%%%%%%%%%%%%%%%%%%%%%%
\begin{table}[ht]
 $$
 \begin{array}{|c|c|c|}
 \hline {\mbox {Parameter} }& NH & IH   \\
 \hline
  \Delta m^2_{sol}/10^{-5} eV^{2}&            6.79 \rightarrow 8.01   &   6.79 \rightarrow 8.01    \\
   \Delta m^2_{atm}/10^{-3} eV^{2}&             2.432 \rightarrow 2.618 &   2.416 \rightarrow  2.603   \\
          \sin^2\theta_{12}&            0.275\rightarrow 0.350 &   0.275 \rightarrow 0.350            \\
          \sin^2\theta_{23}&            0.427 \rightarrow 0.609  &		0.430  \rightarrow  0.612		 \\
          \sin^2\theta_{13}&            0.02046 \rightarrow 0.02440 
 &  	0.02066   \rightarrow 0.02461		\\
 
 \delta_{CP}  &  0.783 \pi  \rightarrow   2.056 \pi & 1.139 \pi \rightarrow  1.967 \pi  \\

   \sin^2\theta_{14} & 0.0098 \rightarrow 0.0310 & 0.0098 \rightarrow 0.0310 \\

   \sin^2\theta_{24} & 0.0059 \rightarrow 0.0262 & 0.0059 \rightarrow 0.0262  \\

   \sin^2\theta_{34} & 0 \rightarrow 0.0396  & 0 \rightarrow 0.0396  \\

   \delta_{14} & 0 \rightarrow 2\pi &  0 \rightarrow 2\pi\\

   \delta_{24} &  0 \rightarrow 2\pi &  0 \rightarrow 2\pi \\
                 
 \hline
 \end{array}
 $$
 \label{table f}\caption{\small{ The oscillation parameters in their $3\sigma$ range,  for NH and IH as given by the global analysis
of neutrino oscillation data with three 
light active neutrinos \cite{Esteban:2018azc} and one extra sterile neutrino \cite{Gariazzo:2017fdh}.
 }}\label{oscdata}\end{table}
%%%%%%%%%%%%%%%%%%%%%%%%%%%%%%%%%%%%%%%%%%%%%%

The neutral current Lagrangian for NSIs in matter is given by the effective dimension 6 four fermion operator as \cite{Wolfenstein:1977ue},
\begin{equation}
\mathcal{L}_{NSI}=-2\sqrt{2}G_F\sum_{f,P,\alpha,\beta} \epsilon^{fP}_{\alpha\beta}(\bar{\nu_\alpha}\gamma^\mu P_L\nu_\beta)(\bar{f}\gamma_\mu Pf),
\label{eq:ak}
\end{equation}
where $f$ is the charged fermion, $P$ is the projection operator (left and right), and $\epsilon^{fP}_{\alpha\beta}$ are the parameters which govern the 
NSIs. The NSI affects the neutrino propagation in matter 
through vector coupling and we can write $\epsilon^{fP}_{\alpha\beta}=\epsilon^{fL}_{\alpha\beta}+\epsilon^{fR}_{\alpha\beta}$. If we assume that the flavour structure of neutrino interaction is independent of charged fermion type, one can write,
\begin{equation}
\epsilon^{fP}_{\alpha\beta}=\epsilon^\eta_{\alpha\beta}\xi^{f,P},
\end{equation}
where $\epsilon^\eta_{\alpha\beta}$ denotes the coupling to the neutrino term and $\xi^{f,P}$ denotes the coupling to the charged fermion term. Hence, Eqn.~(\ref{eq:ak}) can be written as,
\begin{equation}
\mathcal{L}_{NSI}=-2\sqrt{2}G_F\sum_{\alpha,\beta} \epsilon^{\eta}_{\alpha\beta}(\bar{\nu_\alpha}\gamma^\mu P_L\nu_\beta)\sum_{f,P}\xi^{f,P}(\bar{f}\gamma_\mu Pf).
\label{eq:bk}
\end{equation}
It is convenient to write,
\begin{equation}
\epsilon^f_{\alpha\beta}=\epsilon^\eta_{\alpha\beta}\xi^f \hspace{1cm} \rm{with},\hspace{1cm} \xi^f=\xi^{f,L}+\xi^{f,R}.
\end{equation}
We can parametrize the quark coupling in terms of $\eta$ as,
\begin{equation}
\xi^u=\frac{\sqrt{5}}{3}(2\cos\eta-\sin\eta), \hspace{1cm} \xi^d=\frac{\sqrt{5}}{3}(2\sin\eta-\cos\eta).
\end{equation}
The normalization constant is chosen in such a way that $\eta\approx 26.6^{\circ}$ corresponds to $\xi^u=1$ and $\xi^d=0$, which defines NSI with up quark and  $\eta\approx 63.4^{\circ}$ corresponds to $\xi^u=0$ and $\xi^d=1$, which defines NSI with down quark. Under $\eta\rightarrow \eta+\pi$, $\xi^u$ and $\xi^d$ flip sign so it is sufficient to consider the parameter space $-\frac{\pi}{2}\leq\eta \leq\frac{\pi}{2}$. It was shown in \cite{Esteban:2018ppq} that the DLMA solution is allowed at $3\sigma$ range for 
$-38^{\circ}\leq\eta\leq87^{\circ}$. 
They presented the allowed parameter 
space in the $\Delta m^2_{21} - \sin^2\theta_{12}$ plane for 
different values of $\eta$. 
They also gave the allowed range of parameters from a global analysis, 
including NSI, marginalizing over $\eta$. From this analysis, 
the allowed range of 
$\sin^2\theta_{12}$ in presence of NSI is obtained as: 
$sin^2\theta_{12}$:  0.214-0.356 (LMA) and 
$sin^2\theta_{12}$: 0.648 -0.745 (DLMA). 
The allowed range of $\Delta m^2_{21}$ varies in the range  
$(6.73  - 8.14) \times 10^{-5}$ eV$^2$  for the LMA solution 
and $(6.82 - 8.02) \times 10^{-5}$ eV$^2$ for the DLMA solution. 
Comparing with the values given in Table~\ref{oscdata}, 
we can see that the range of $\sin^2\theta_{12}$ and $\Delta m^2_{21}$ 
for  LMA solution in presence of 
NSI  have changed only marginally. The other three generation parameters were shown to be stable with inclusion of NSI \cite{Esteban:2018ppq}. 
Thus, in our analysis we use the values of the parameters from Table~\ref{oscdata}
excepting for the parameters $\sin^2\theta_{12}$ and $\Delta m^2_{21}$ 
for which we use the values quoted above from global analysis 
performed in  \cite{Esteban:2018ppq}.  
%\footnote{In a more recent analysis \cite{Coloma:2019mbs}, 
% it was shown that inclsion of COHERENT data constrains the range of $\eta$ for 
%which DLMA solution is allowed even further.
%Including COHERENT total rate the allowed value of $\eta$ is 
%$-38^{\circ}\leq\eta\leq15^{\circ}$ and including the timing and energy information in COHERENT the allowed range of $\eta$ is further restricted as 
%$-38^{\circ}\leq\eta\leq-18^{\circ}$. However, in \cite{ } the allowed ranges of oscillation parameters are not mentioned and hence we take these from 
%the earlier analysis \cite{Esteban:2018ppq}.}  
%}

% In Table-\ref{tableI}, we have chosen the parameter space for $\sin^2\theta_{12}$ corresponding to both LMA and DLMA solution obtained from global oscillation analysis data at $3\sigma$ \cite{Esteban:2018ppq}. In Table \ref{oscdata}, all the neutrino oscillation parameters except $\sin^2\theta_{12}$ are robust in their $3\sigma$ range.} 

The total matter potential including standard and non-standard 
interactions  is governed by the Hamiltonian,
\begin{equation}
H^{sterile+NSI}_{mat}=\sqrt{2} G_F N_e(r)\begin{bmatrix}
1 & 0 & 0 & 0\\
0 & 0 & 0 & 0\\
0 & 0 & 0 & 0\\
0 & 0 & 0 & 0
\end{bmatrix}+\frac{G_F N_n}{\sqrt{2}}\begin{bmatrix}
0 & 0 & 0 & 0\\
0 & 0 & 0 & 0\\
0 & 0 & 0 & 0\\
0 & 0 & 0 & 1
\end{bmatrix}+\sqrt{2} G_F\sum_{f=e,u,d} N_f(r)\begin{bmatrix}
\epsilon^f_{ee} & \epsilon^f_{e\mu} & \epsilon^f_{e\tau} & 0\\
\epsilon^{f*}_{e\mu} & \epsilon^f_{\mu\mu} & \epsilon^f_{\mu\tau} & 0\\
\epsilon^{f*}_{e\tau} & \epsilon^{f*}_{\mu\tau} & \epsilon^f_{\tau\tau} & 0\\
0 & 0 & 0 & 0
\end{bmatrix},
\end{equation}
where, $N_e,\, N_n $ and $N_f$ are the number densities of electron, neutron and the fermion $f$ in the sun. Here, we have neglected non-standard interactions in the sterile sector \footnote{Studies including non-standard interactions of sterile neutrinos have been discussed in \cite{Esmaili:2018qzu}.}. We can now construct the Hamiltonian in an effective $2\times 2$ model as $H^{eff}=H^{eff}_{vac}+H^{eff}_{mat}$ where,
\begin{equation}
H^{eff}=\frac{\Delta m^2_{21}}{4E}\begin{bmatrix}
-\cos2\theta_{12} & \sin2\theta_{12}\\
\sin2\theta_{12} & \cos2\theta_{12}
\end{bmatrix}+A_i \begin{bmatrix}
 c^2_{13}c^2_{14} & 0\\
0 & 0
\end{bmatrix}+A_j\begin{bmatrix}
-k_1 & k_2\\
k^*_2 & k_1
\end{bmatrix}+
A_i\sum_{f=e,u,d} \frac{N_f}{N_e}\begin{bmatrix}
-\epsilon^f_D & \epsilon^f_N\\
\epsilon^{f*}_N & \epsilon^f_D
\end{bmatrix}.\label{H_2x2}
\end{equation}

Here, $A_i=\sqrt{2}G_F N_e$, $A_j=\frac{G_F N_n}{\sqrt{2}}$
and 
we have taken $\theta_{34}=0$. Now the new parameters $\epsilon^f_D, \epsilon^f_N$ are related to the old parameters $\epsilon^f_{\alpha\beta}$ through the following equations :
\begin{equation}
\begin{split}
\epsilon^f_D=c_{13}s_{13}Re[e^{i\delta_{CP}}(s_{23}c_{24}c_{14}\epsilon^f_{e\mu}+c_{14}c_{23}\epsilon^f_{e\tau})]-(1+s^2_{13}) c_{23}s_{23}c_{24}Re(\epsilon^f_{\mu\tau})\\-\frac{c^2_{13}}{2}(\epsilon^f_{ee}c^2_{14}-\epsilon^f_{\mu\mu} c^2_{24})+\frac{s^2_{23}-s^2_{13}c^2_{23}}{2}(\epsilon^f_{\tau\tau}-c^2_{24}\epsilon^f_{\mu\mu})+ c^2_{13} c_{14}s_{14} s_{24} Re(\epsilon^f_{e\mu}e^{i(\delta_{14}-\delta_{24}})\\-c_{13}c_{23} s_{14} s_{24} s_{13} Re(\epsilon^f_{\mu\tau}e^{i(\delta_{CP}-\delta_{14}+\delta_{24})})- \epsilon^f_{\mu\mu}s_{13}c_{24}s_{23}c_{13}s_{14}s_{24}Re(e^{i(\delta_{CP}-\delta_{14}+ \delta_{24})})\\ -\frac{\epsilon^f_{\mu\mu}}{2}s^2_{14}s^2_{24}c^2_{13}
\end{split}
\end{equation}
and
\begin{equation}
\begin{split}
\epsilon^f_N=c_{13}[c_{14}c_{24}c_{23}\epsilon^f_{e\mu}-c_{14}s_{23}\epsilon^f_{e\tau}]+s_{13}e^{-i\delta_{CP}}[\epsilon^f_{\mu\tau}s^2_{23}c_{24}-c^2_{23}c_{24}\epsilon^{f*}_{\mu\tau}\\+c_{23}s_{23}(\epsilon^f_{\tau\tau}-\epsilon^f_{\mu\mu}c^2_{24})]+e^{-i(\delta_{14}-\delta_{24})}c_{13}s_{14}s_{24}(\epsilon^f_{\mu\tau}s_{23}-\epsilon^f_{\mu\mu}c_{23}c_{24})
\end{split}.
\end{equation}

$k_1$ and $k_2$ are defined as,
\begin{equation}
k_1=\frac{1}{2}(c^2_{23}s^2_{24}-c^2_{13}c^2_{24}s^2_{14}-s^2_{13}s^2_{23}s^2_{24})+s_{13}s_{23}s_{24}c_{13}c_{24}s_{14}Re(\delta_{14}-\delta_{CP}-\delta_{24}),
\end{equation}
\begin{equation}
k_2=e^{i(\delta_{24}-\delta_{14})}c_{23}s_{24}c_{13}c_{24}s_{14}-e^{-i\delta_{CP}}s_{13}s_{23}s^2_{24}c_{23}.
\end{equation}
In the absence of sterile neutrino ( $\theta_{i4}=0$ and $\delta_{i4}=0$ where $i=1,2$ )implying $k_1=k_2=0$,  we get back the expressions of $\epsilon^f_D$ and $\epsilon^f_N$ of \cite{Gonzalez-Garcia:2013usa}.\\
%%%%%%%%%%%%%%%%%%%%%%%%%%%%%%%%%
Now we define $\delta=\frac{\Delta m^2_{21}}{2E}$, 
$\alpha_f=\frac{N_f}{N_e}$, 
%$A_j=\frac{G_F N_n}{\sqrt{2}}$ i
and rewrite Eqn.~(\ref{H_2x2}) as,
%%%%%%%%%%%%%%%%%%%%%%%%%%%%%%%%%%%%
%\begin{equation}
%H_{eff}=H^{eff}_{vac}+H^{eff}_{mat}
%\end{equation}
%%%%%%%%%%%%%%%%%%%%%%%%%%%%%%%%%%%%
\begin{equation} H^{eff}
=\frac{\delta}{2}\begin{bmatrix}
-\cos2\theta_{12} & \sin2\theta_{12}\\
\sin 2\theta_{12} & \cos2\theta_{12}
\end{bmatrix}+ A_i \begin{bmatrix}
c^2_{13}c^2_{14} & 0\\
0 & 0
\end{bmatrix}+A_j\begin{bmatrix}
-k_1 & k_2\\
k^*_2 & k_1
\end{bmatrix}+A_i\sum_{f=e,u,d} \alpha_f\begin{bmatrix}
-\epsilon^f_D & \epsilon ^f_N\\
\epsilon^{f*}_N & \epsilon^f_D
\end{bmatrix}.
\end{equation}
Diagonalizing the above effective Hamiltonian gives the matter mixing angle $\theta_M$ as,
\begin{equation}
\tan2\theta_M=\frac{\delta\sin2\theta_{12}+2A_i\alpha_f\epsilon^f_N+2A_j k_2}{\delta\cos2\theta_{12}+2A_i\alpha_f\epsilon^f_D-A_ic^2_{13}c^2_{14}+2A_j k_1}.
\end{equation}

Hence, the resonance occurs when,
\begin{equation}
\delta\cos2\theta_{12}+2\alpha_fA_i\epsilon^f_D=A_ic^2_{13}c^2_{14}-2A_j k_1,
\end{equation}
i.e., \begin{equation}
\Delta m^2_{21}\cos2\theta_{12}+Bk_1=A[c^2_{13}c^2_{14}-2\alpha_f\epsilon^f_D]
\label{solar-res}.
\end{equation}
Here, $A= 2\sqrt{2}G_F N_e E$ and $B=  2\sqrt{2}G_F N_n E$.

\begin{figure}[tbh]
    \includegraphics[scale=0.4]{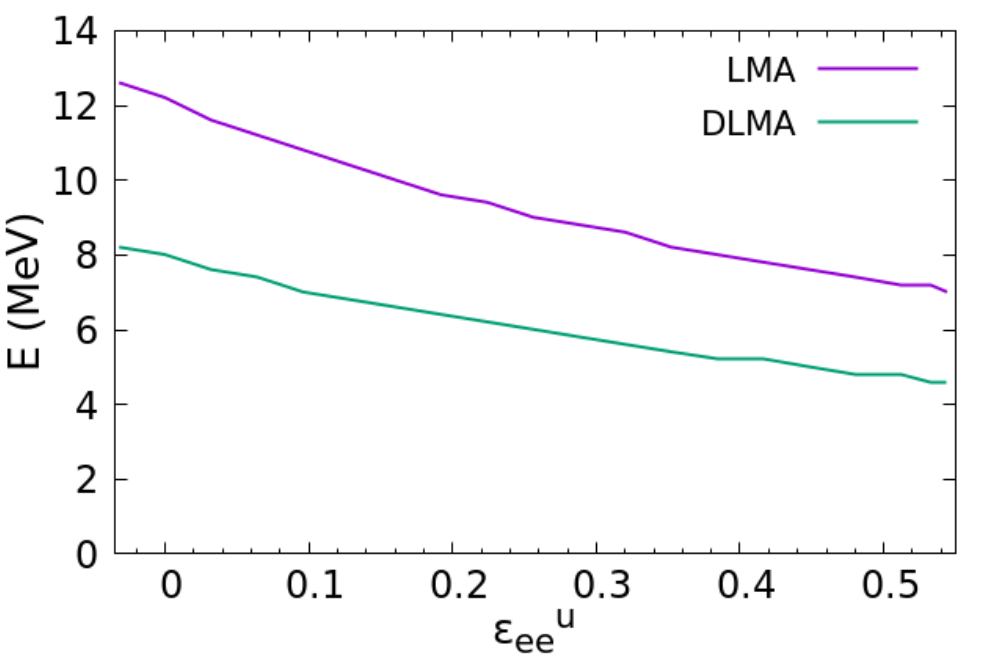}
    \caption{  The energies corresponding to resonance for different values of $\epsilon^u_{ee}$ for LMA (purple line) and DLMA (green line) solutions.} \label{eee}
\end{figure}

It is crucial to ensure  the occurrence of solar neutrino resonance with DLMA solution in a 3+1 neutrino scenario before we proceed to study the implications in $0\nu \beta\beta$ process. Keeping this in mind, we have used the resonance condition in Eqn.~(\ref{solar-res}) and obtained the neutrino energies at which the solar neutrino resonance occurs.  For this study we have only considered $\epsilon_{ee}^u$ to be non-zero while setting other NSI parameters to be 0 for simplicity. In Fig.\ref{eee}, we have plotted the energy for which MSW resonance occurs for different values of $\epsilon_{ee}$ for both LMA (purple line) and the DLMA (green line) solutions. The figure shows that for $\sin^2\theta_{12}$ in the DLMA region, resonance 
condition can be obtained for different values of $\epsilon_{ee}^{u}$, but for 
a lower energy. The chosen values of $\epsilon_{ee}^{u}$ are within the range allowed by the global analysis of  data as given in reference
\cite{Esteban:2018ppq}.
This is a preliminary verification and a detailed analysis is beyond the scope of this paper.

%%%%%%%%%%%%%%%%%%%%%%%%%%%%%%%%%%%%
%%%%%%%%%%%%%%%%%%%%%%%%%%%%%%%%%%%%

\section{$0\nu\beta\beta$ in 3+1 scenario}\label{section3}

The half life for $0\nu\beta\beta$ in the standard scenario with 
light neutrino exchange is given by \cite{Doi:1985dx,Haxton:1985am},
\begin{equation}
(T_{1/2})^{-1}=G\Big|\frac{M_\nu}{m_e}\Big|^2 m^2_{\beta\beta},\label{0nbbthalf}
\end{equation}
where $G$ is the phase space factor, $M_\nu$ is the nuclear matrix element and $m_e$ is the electron mass. The expression for the effective Majorana mass $m_{\beta\beta}$ is given by,
\begin{equation}
m_{\beta\beta}=|U^2_{ei}m_i|,\label{meff}
\end{equation}
where $i$ runs from $1$ to $3~(4)$ in the case of three (four) generations. 
$m_i$ denotes the mass eigenstates and $U$ is the unitary PMNS matrix 
as given in Eqn.~\ref{eq:pmns}.

Thus, in 3+1 scheme,
\begin{equation}
m_{\beta\beta}=|m_1c^2_{12}c^2_{13}c^2_{14}+m_2 s^2_{12}c^2_{13}c^2_{14}e^{i\alpha}+m_3s^2_{13}c^2_{14}e^{i\beta} + m_4 s^2_{14}e^{i\gamma}|,
\label{eq:mb}
\end{equation}
where we have used the usual convention with $c_{ij}=\cos\theta_{ij}$ and
$s_{ij}=\sin\theta_{ij}$. The above expression for $m_{\beta\beta}$ in the case of four generation is related to that in the case of three generation as,
\be  {m_{\beta\beta}}_{4gen} = |c_{14}^2~ {m_{\beta\beta}}_{3gen} + m_4 s^2_{14}e^{i\gamma}  | .\ee

Thus, the $m_{\beta\beta}$ in the case of four generation depends on three extra parameters : the mixing angle $\theta_{14}$, the third mass squared difference $\Delta m_{LSND}^2$ ($m_4 = \sqrt{m_1^2 + \Delta m_{LSND}^2 }$) and the Majorana phase $\gamma/2$. Depending on the values of these parameters, there can be additional enhancement or cancellation in the predictions of $m_{\beta\beta}$ compared to that in the three generation case. In this work, we denote the standard LMA solution as $\theta_{12}$ and the DLMA solution as $\theta_{D12}$. The $3\sigma$ ranges of these two parameters in the presence of NSI are shown in Table-\ref{tableI} \cite{Esteban:2018ppq}.

%%%%%%%%%%%%%%%%%%%%%%%%%%%%%%%%%%%%%%%%%%%%%%%%%%%%%%%%%%%%%%
\begin{table}[ht]
 $$
 \begin{array}{|c|c|c|c|c|c|}
 \hline {\mbox {} }& \textrm{sin}^2\theta_{12} & \textrm{sin}^2\theta_{D12} & \textrm{cos}2\theta_{12}    &   \textrm{cos}2\theta_{D12}   &   \textrm{sin}^2\theta_{13}  \\
 \hline
 
  Maximum  & 0.356  & 0.745 &  0.57   & -0.296   &    0.024 \\
  
  \hline
 
  Minimum  & 0.214  &   0.648   &  0.29   & -0.49     &   0.020  \\
   
 \hline
 \end{array}
 $$
\caption{\small{ The $3\sigma$ ranges of 
different combinations of oscillation parameters in the presence of NSI 
relevant for understanding the behavior of the effective mass in different limits\cite{Esteban:2018ppq}. 
 }}\label{tableI}\end{table}

%%%%%%%%%%%%%%%%%%%%%%%%%%%%%%%%%%%%%%%%%%%%%%%%%%%%%%%%%%%%%%%
$m_{\beta\beta}$ is highly sensitive to the mass hierarchy of the light neutrinos, i.e; whether $m_1$ or $m_3$ is the lowest mass eigenstate.\\For normal hierarchy (NH), $m_1$ is the lowest mass eigenstate $(m_1<m_2<<m_3)$ and we can express the other mass eigenstates in terms of $m_1$ as
\begin{equation}
m_2=\sqrt{m^2_1+\Delta m^2_{sol}}\hspace{1cm} m_3=\sqrt{m^2_1+\Delta m^2_{atm}}\hspace{1cm} m_4=\sqrt{m^2_1+\Delta m^2_{LSND}}.
\end{equation}
For inverted hierarchy (IH), $m_3$ is the lowest mass eigenstate $(m_3<<m_1\approx m_2)$ and the other mass eigenstates in terms of $m_3$ are,
\begin{equation}
m_1=\sqrt{m^2_3+\Delta m^2_{atm}}\hspace{1cm} m_2=\sqrt{m^2_3+\Delta m^2_{sol}+\Delta m^2_{atm}} \hspace{1cm} m_4=\sqrt{m^2_3+\Delta m^2_{atm}+\Delta m^2_{LSND}}.
\end{equation}
Here, $\Delta m^2_{sol}=m^2_2-m^2_1$, $\Delta m^2_{atm}=m^2_3-m^2_1 (m^2_1-m^2_3)$ for NH(IH) and $\Delta m^2_{LSND}=m^2_4-m^2_1$.

%%%%%%%%%%%%%%%%%%%%%%%%%%%%%%%%%%%%%%%%%%%%
\begin{figure}
\includegraphics[width=8.5cm]{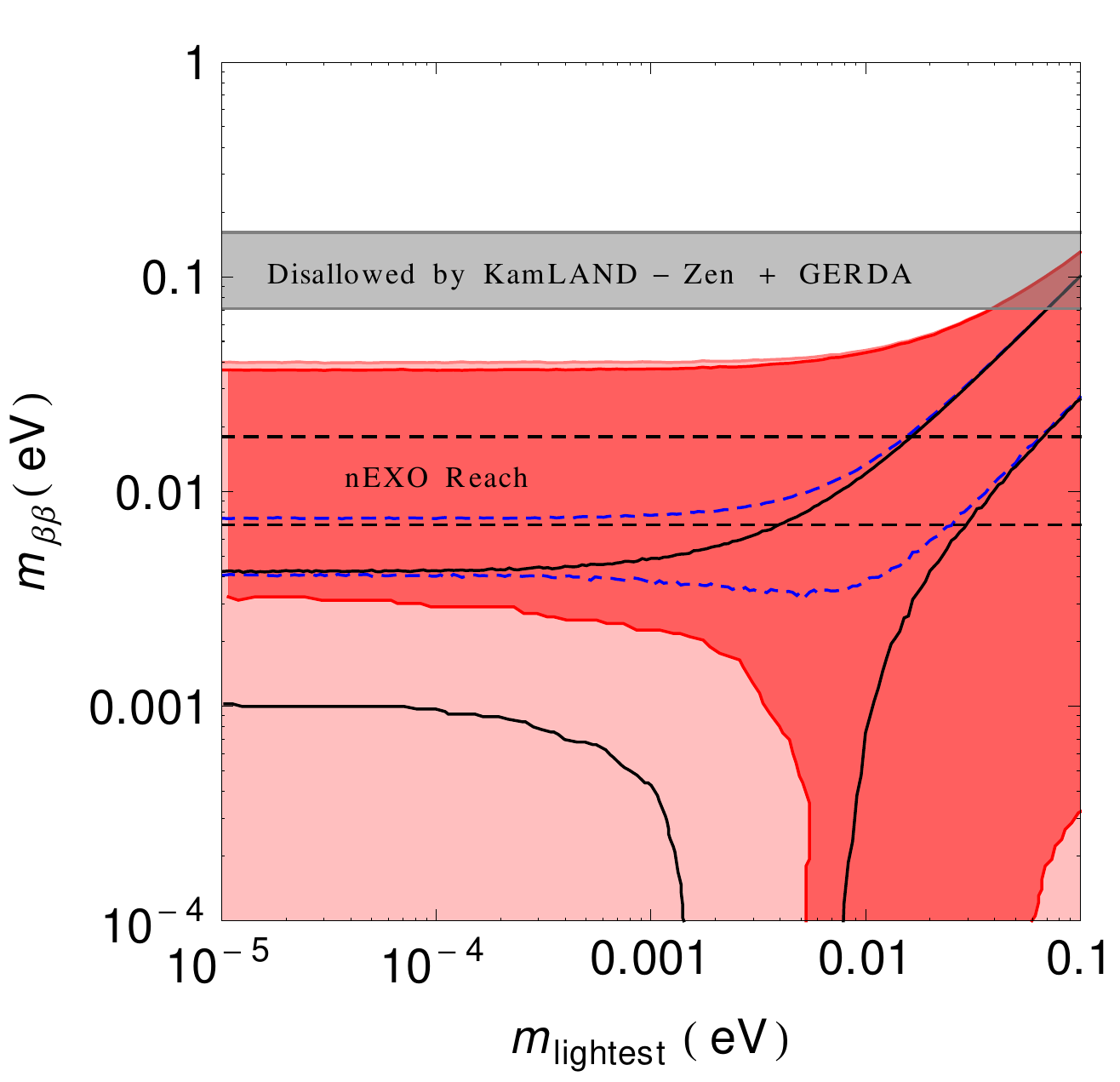} 
\includegraphics[width=8.5cm]{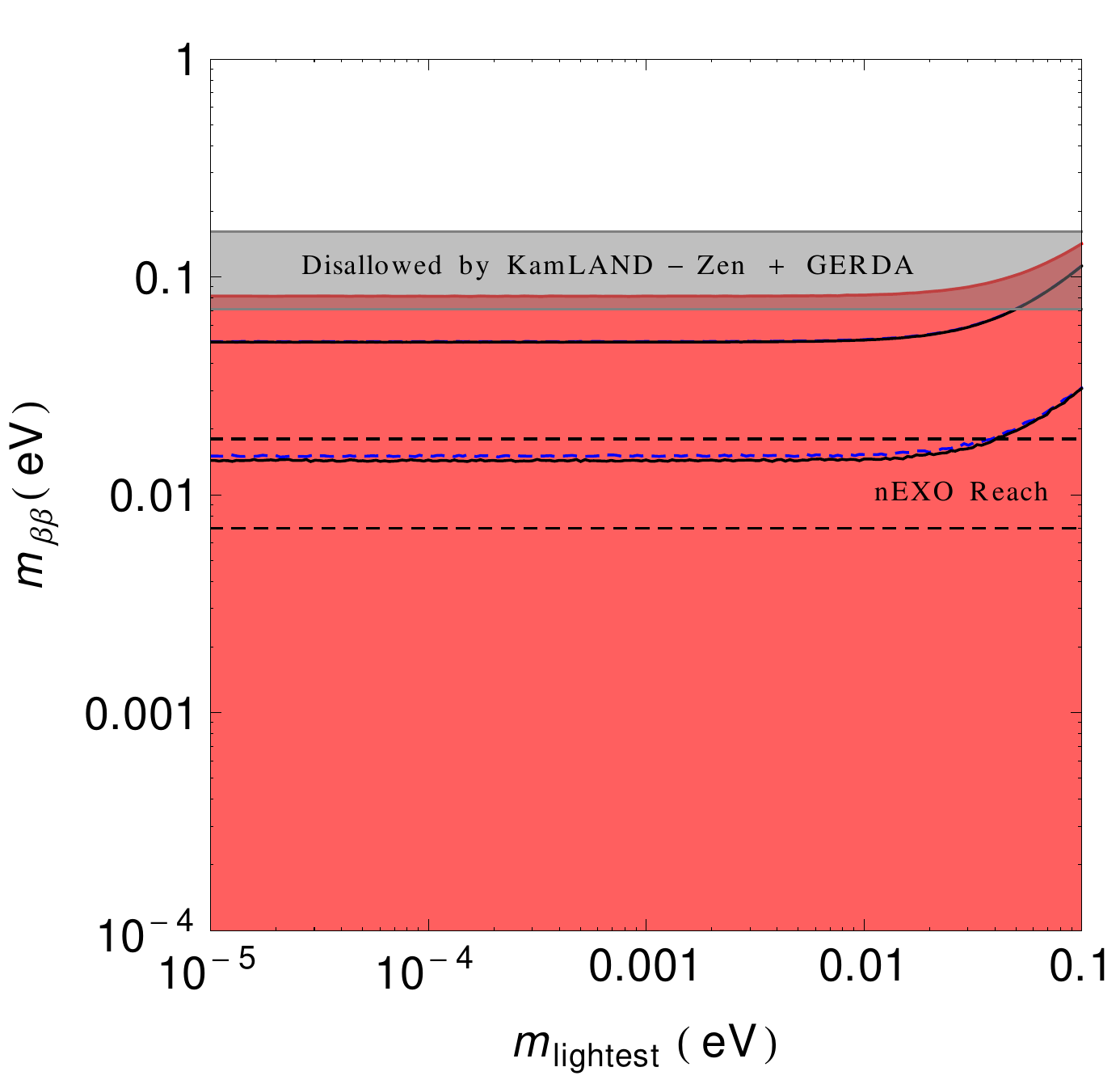} 
\caption{$m_{\beta\beta}$ vs $m_{lightest}$ for NH (left) and IH (right) for $\Delta m^2_{LSND}=1.3 \textrm{eV}^2$. The pink and the red regions represent the predictions for the standard LMA as well as the DLMA solutions for $\theta_{12}$ respectively. The gray shaded region represents the current upper bound of $m_{\beta\beta}$ obtained from the combined results of KamLAND-Zen and GERDA experiments and the band defined by the two horizontal black dashed lines represents the future $3\sigma$ sensitivity of the 
nEXO experiment. The black solid lines and the blue dotted lines represent the predictions with the standard three neutrino case for the standard LMA and the DLMA solutions respectively.}
\label{fig:er}
\end{figure}
%%%%%%%%%%%%%%%%%%%%%%%%%%%%%%%%%%%%%%%%%%%%%%%%

%%%%%%%%%%%%%%%%%%%%%%%%%%%%%%%%%%%%%%%%%%%%%%%%%
\begin{figure}
\includegraphics[width=8.5cm]{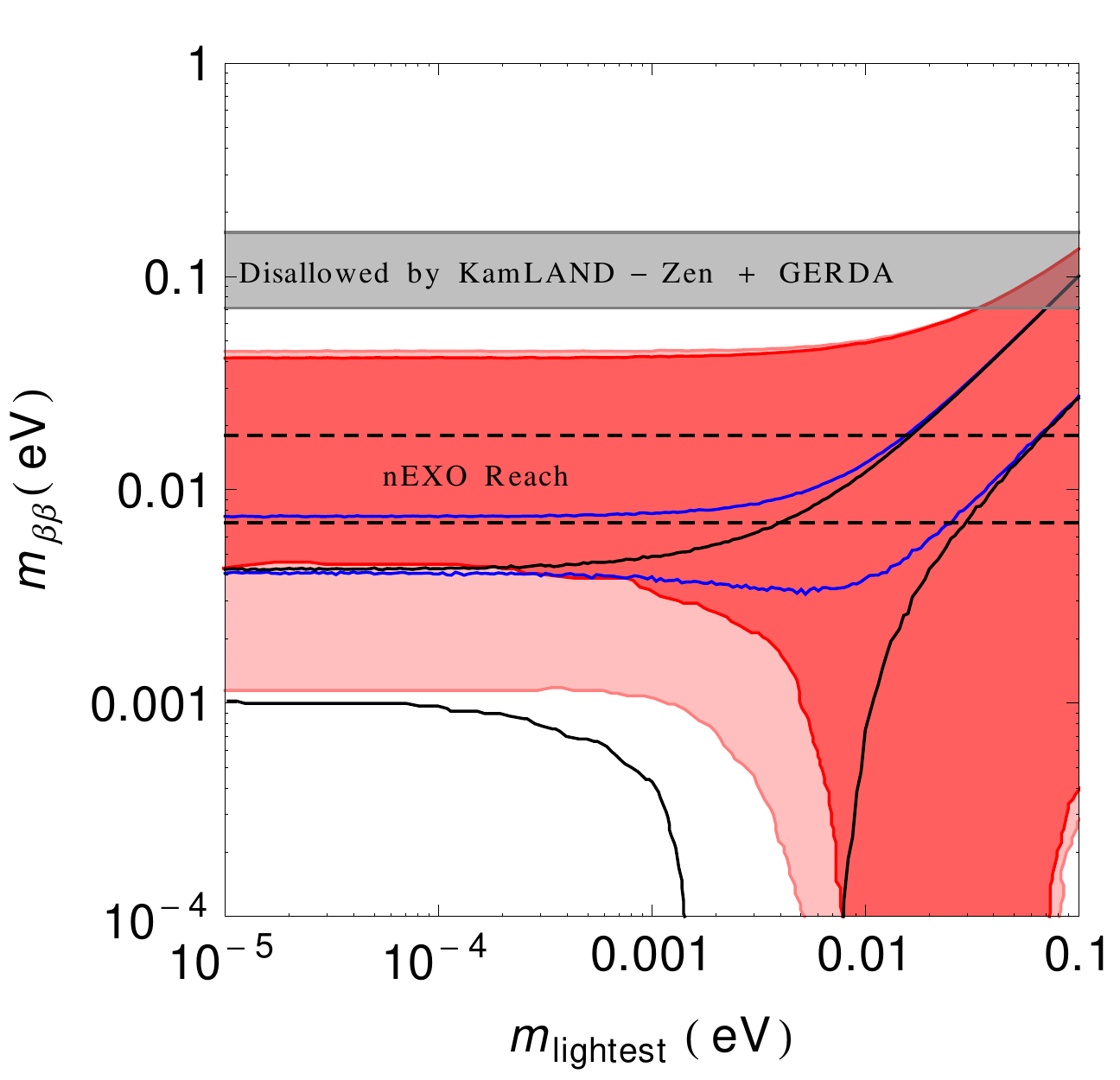} 
\includegraphics[width=8.5cm]{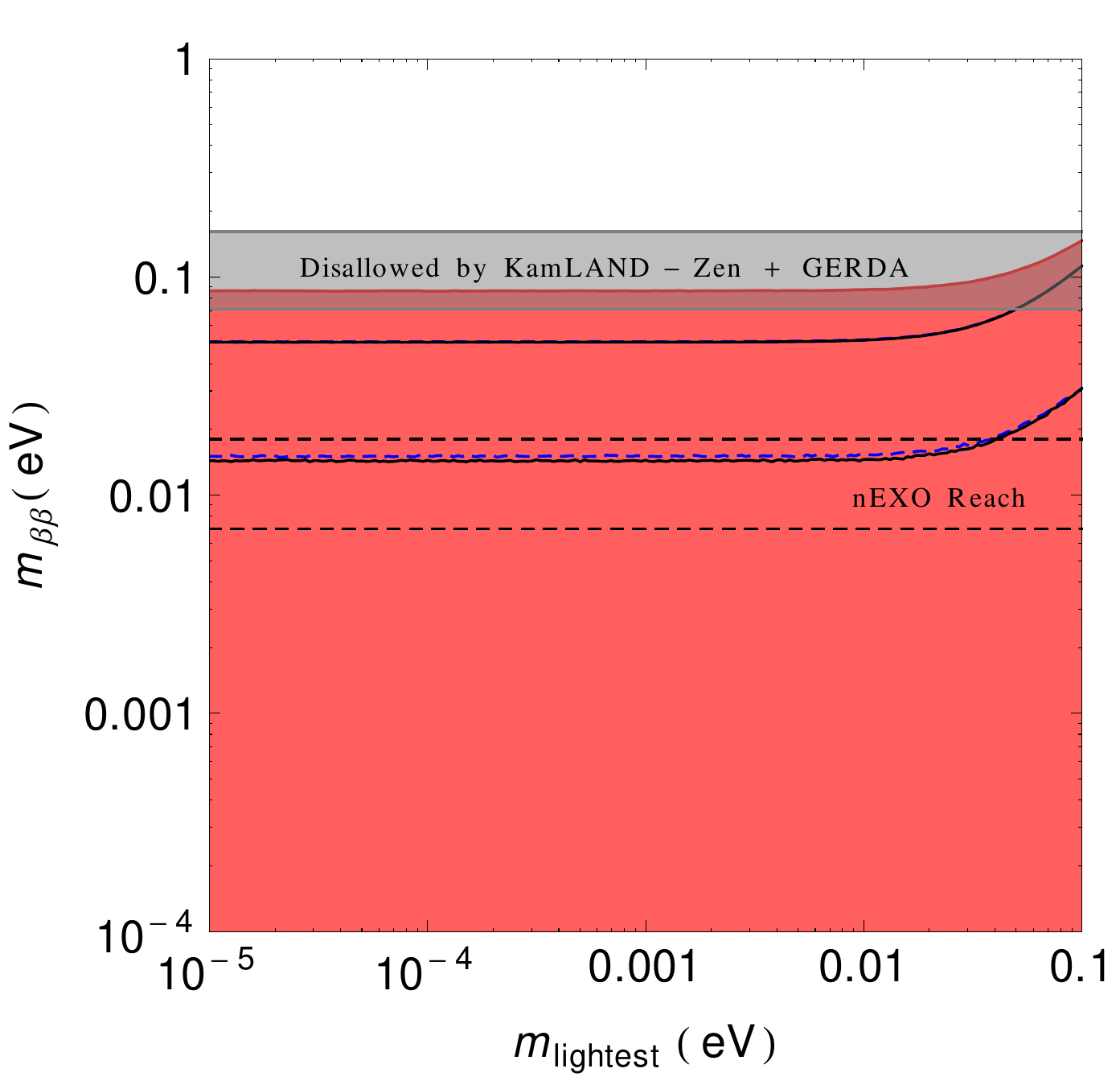} 
\caption{$m_{\beta\beta}$ vs $m_{lightest}$ for NH (left) and IH (right) for $\Delta m^2_{LSND}=1.7 \textrm{eV}^2$. The pink and the red regions represent the predictions for the standard LMA as well as the DLMA solutions for $\theta_{12}$ respectively. The gray shaded region represents the current upper bound of $m_{\beta\beta}$ obtained from the combined results of KamLAND-Zen and GERDA experiments and the band defined by the two horizontal black dashed lines represents the future $3\sigma$ sensitivity of the 
nEXO experiment. The black solid lines and the blue dotted lines represent the predictions with the standard three neutrino case for the standard LMA and the DLMA solutions respectively.}
\label{fig:es}
\end{figure}
%%%%%%%%%%%%%%%%%%%%%%%%%%%%%%%%%%%%%%%%%%%%%%%%%

In Figs.~\ref{fig:er} and \ref{fig:es} we have shown the predictions for $m_{\beta\beta}$ as a function of the lightest neutrino mass for two different values of the third mass squared difference, i.e., $\Delta m^2_{LSND}=1.3\textrm{eV}^2$ and $1.7 \textrm{eV}^2$. The left panels are for NH whereas the right panels are for IH. In plotting these figures, we have varied the oscillation parameters in their $3\sigma$ ranges \cite{Esteban:2018azc, Esteban:2018ppq}, the Majorana phases in the range $0-\pi$ and the mixing angle $\theta_{14}$ in the range $\theta_{14} \sim 0.08 - 0.17$ radian.

In these plots, the pink and the red regions represent the predictions for the standard LMA as well as the DLMA solutions for $\theta_{12}$ respectively. The gray shaded region in the range between $0.071~eV$ and $0.161~eV$ represents the current upper bound of $m_{\beta\beta}$ obtained from the combined results of KamLAND-Zen and GERDA experiments \cite{Agostini:2018tnm}. This is a band due to the NME uncertainties \cite{Engel:2016xgb,Agostini:2018tnm,Kotila:2012zza}. The region above this band is disallowed. The band defined by the two horizontal black dashed lines represents the future $3\sigma$ sensitivity of the 
nEXO experiment : $T_{1/2} = 5.7 \times 10^{27}$ years \cite{Kharusi:2018eqi}, which, has been converted to $m_{\beta\beta} = 0.007-0.018$ eV using Eqn.~\ref{0nbbthalf} by including the NME uncertainties. The black solid lines and the blue dotted lines represent the predictions for $m_{\beta\beta}$ with the standard three neutrino case for the standard LMA and the DLMA solutions respectively \cite{N.:2019cot}.

From Figs.~\ref{fig:er} and \ref{fig:es}, we can see that in the case of IH, the predictions of $m_{\beta\beta}$ remains same for both LMA and DLMA solutions and this is true for both the three generation as well as four generation cases. In addition, these predictions are independent of the values of $\Delta m_{LSND}^2$ that we have considered. Also, complete cancellation of $m_{\beta\beta}$ can occur for the entire range of $m_3$ in the presence of the fourth sterile 
neutrino, unlike in the three generation case where there is no cancellation 
region for IH at all. In addition, the maximum predicted values for
$m_{\beta\beta}$ are higher in the case of the four generation. 
Also, one can see that even though the non-observation of a positive signal for $0\nu\beta\beta$ in the future nEXO experiment will rule out the IH scenario in the case of three generation, it can still be allowed in the presence of the 
fourth sterile neutrino for both LMA and DLMA solution. 
In fact, the maximum value of $m_{\beta\beta}$ in this case is already in the 
region disallowed by the present results on $0\nu\beta\beta$, subject to
the NME uncertainty. This can be 
used to constrain the $\theta_{14}$ mixing angle \cite{Chakraborty:2019rjc} .

 In the case of NH, complete cancellation can occur for certain values of
 $m_1$ for both the standard LMA as well as the DLMA solutions in the four generation case, whereas for the three generation case, there is no cancellation region for the DLMA solution. The values of $m_{lightest}$ for which complete cancellation of $m_{\beta\beta}$ occurs is larger for the DLMA solution.
There is more cancellation region for $\Delta m_{LSND}^2=1.3~ \textrm{eV}^2$ compared to that for $\Delta m_{LSND}^2=1.7~ \textrm{eV}^2$. For $\Delta m_{LSND}^2=1.3~ \textrm{eV}^2$ with the standard LMA solution, cancellation is possible in the entire range of $m_{lightest}$ as in the case of IH.
But for the DLMA solution cancellation is possible only for higher values 
of $m_{lightest}$. 
Another important point to be noted is that for the 
sterile neutrino scenario, there is no desert region between 
NH and IH unlike in the standard three generation picture \cite{N.:2019cot}. 
This is true for both LMA and DLMA solutions. 
Also, the maximum allowed values of $m_{\beta\beta}$ is higher in the case of 
the four generation picture and is almost independent of whether one take the standard LMA or the DLMA solution. 
However, as compared to the three generation DLMA, the predictions for the 
maximum value of  $m_{\beta \beta}$ are higher for the sterile neutrino case. 
The prediction of $m_{\beta \beta}$ for three neutrino DLMA picture 
is in the range (0.004-0.0075) eV   
while for the sterile DLMA(and LMA) this spans (0.004 - .04) eV (for $m_{lightest} \lesssim 0.005$ eV) for NH. 
The new allowed region of 0.0075-0.04 eV in the case of NH with four generation 
is in the complete reach of the future nEXO experiment.

The behavior of the effective Majorana mass  $m_{\beta\beta}$ for the two different mass orderings can be understood by considering various limiting cases.\\
\begin{itemize}
\item Inverted Hierarchy:
We discuss the following limiting cases: \\

Case I : For $m_3<<\sqrt{\Delta m^2_{atm}}, m_1\approx m_2\approx \sqrt{\Delta m^2_{atm}}$ and $m_4=\sqrt{\Delta m^2_{LSND}}$ the effective mass parameter from  Eqn.~\ref{eq:mb} becomes,
\begin{equation}
m_{\beta\beta IO}\approx|\sqrt{\Delta m^2_{atm}} c^2_{13}c^2_{14}(c^2_{12}+s^2_{12}e^{i\alpha})+\sqrt{\Delta m^2_{LSND}}s^2_{14}e^{i\gamma}|.
\label{eq:ih}
\end{equation}
Here we take the representative  values of $\Delta m^2_{atm}=2.5\times 10^{-3}\textrm{eV}^2$ and $\Delta m^2_{LSND}=1.3\textrm{eV}^2$. The above equation can lead to cancellation if we choose the following approximations $c^2_{13}\sim c^2_{14}\sim 1, s^2_{12}\sim 0.356, c^2_{12}\sim 0.644, \sqrt{\Delta m^2_{atm}}\sim 0.05, \sqrt{\Delta m^2_{LSND}}\sim \sqrt{1.3}\sim 1.140$. This implies,
\begin{equation}
m_{\beta\beta}=0.0322+0.0178 e^{i\alpha}+1.14 s^2_{14}e^{i\gamma}.
\end{equation} 
So the cancellation region corresponds to $\alpha\sim \pi$, $\gamma\sim \pi$ and $s^2_{14}\sim 0.0126$. The cancellation is achieved due to large value of  $\sqrt{\Delta m^2_{LSND}}$. In three generation case, such cancellation is not there because of the absence of large valued term which can counter the first large positive  term. In this region, the effective mass parameter is independent of the lightest neutrino mass eigenstate (Eqn.~\ref{eq:ih}) and is bounded from above and below by,
\begin{equation}
m_{\beta\beta IO max}=|\sqrt{\Delta m^2_{atm}} c^2_{13}c^2_{14}+\sqrt{\Delta m^2_{LSND}}s^2_{14}| ~;~ (\alpha=0,2\pi; \gamma=0,2\pi).
\end{equation} 
\begin{equation}
m_{\beta\beta IO min}=|\sqrt{\Delta m^2_{atm}} c^2_{13}c^2_{14}\cos2\theta_{12}-\sqrt{\Delta m^2_{LSND}}s^2_{14}|~;~ (\alpha=\pi; \gamma=\pi).
\end{equation}

The maximum value of $m_{\beta\beta}$ is independent of $\theta_{12}$ whereas the minimum value of $m_{\beta\beta}$ depends on $\theta_{12}$. But the minimum value of $m_{\beta\beta}$ is of the order of $\sim 10^{-4}$ and hence, this difference is not much pronounced.

Case II : As $m_3$ approaches to $\sqrt{\Delta m^2_{atm}}$, the other mass states are $m_1\approx m_2\approx \sqrt{2\Delta m^2_{atm}}$, $m_4\approx\sqrt{\Delta m^2_{LSND}}$ and $m_{\beta\beta}$ from Eqn.~\ref{eq:mb} becomes,
\begin{equation}
m_{\beta\beta}=|\sqrt{2\Delta m^2_{atm}}c^2_{13}c^2_{14}(c^2_{12}+s^2_{12}e^{i\alpha})+\sqrt{\Delta m^2_{atm}}s^2_{13}c^2_{14}e^{i\beta}+\sqrt{\Delta m^2_{LSND}}s^2_{14}e^{i\gamma}|.
\label{eq:rh}
\end{equation}
Using the same values of those parameters as in case I and $s^2_{13}\sim 0.024$ we have,
\begin{equation}
m_{\beta\beta}=0.0455+0.0252 e^{i\alpha}+1.2\times 10^{-3}e^{i\beta}+1.14 s^2_{14}e^{i\gamma}.
\end{equation}
Here the cancellation occurs for $\alpha\sim\beta\sim\gamma\sim \pi$ and $s^2_{14}\sim 0.017$. In three neutrino mixing the cancellation is not possible due to the absence of large $m_4$ term. Eqn.~\ref{eq:rh} is maximum for $\alpha,\beta,\gamma=0,2\pi$ and is independent of $\theta_{12}$. Since the value of $s^2_{13}$ is small, the minimum value of $m_{\beta\beta}$ is independent of $\beta$. Hence the minimum value of $m_{\beta\beta}$ corresponds to $\alpha\sim\pi$ and $\gamma\sim\pi$, and is given as,
\begin{equation}
m_{\beta\beta IO min}=|\sqrt{2\Delta m^2_{atm}}c^2_{13}c^2_{14}\cos2\theta_{12}-\sqrt{\Delta m^2_{LSND}}s^2_{14}|.
\end{equation}

\item Normal Hierarchy:
We consider the following 
limiting cases:   \\
Case I: If $m_1<<m_2\approx \sqrt{\Delta m^2_{sol}}<<m_3\approx\sqrt{\Delta m^2_{atm}}$, then $m_{\beta\beta}$ can be written from Eqn.~\ref{eq:mb} as,
\begin{equation}
m_{\beta\beta}=\sqrt{\Delta m^2_{atm}}\Big|s^2_{13}c^2_{14}e^{i\beta}+\frac{\sqrt{\Delta m^2_{sol}}}{\sqrt{\Delta m^2_{atm}}}s^2_{12}c^2_{13}c^2_{14}e^{i\alpha}+\frac{\sqrt{\Delta m^2_{LSND}}}{\sqrt{\Delta m^2_{atm}}}s^2_{14}e^{i\gamma}\Big|.
\end{equation}
Taking the same representative values as we have used in the discussion for IH, we have,
\begin{equation}
m_{\beta\beta}=\sqrt{\Delta m^2_{atm}}|s^2_{13}c^2_{14}e^{i\beta}+0.172s^2_{12}c^2_{13}c^2_{14}e^{i\alpha}+22.80s^2_{14}e^{i\gamma}|,
\end{equation}
or,
\begin{equation}
m_{\beta\beta}=\sqrt{\Delta m^2_{atm}}|0.024 e^{i\beta}+0.061 e^{i\alpha}+22.80 s^2_{14}e^{i\gamma}|.
\end{equation}
We take the value of $\sin\theta_{14}$ in the range $0.08$ to $0.17$, which implies for small $m_1$ there is no cancellation since the value of $m_4$ is very large. The maximum value of $m_{\beta\beta}$ corresponds to $\alpha,\beta,\gamma=0,2\pi$ and the minimum value corresponds to $\gamma=0$ and $\alpha,\beta=\pi$. $m_{\beta\beta}$ is higher for higher value of $\sin^2\theta_{12}$. This implies that $m_{\beta\beta}$ for the DLMA solution is higher in this region.

Case II: As $m_1\sim\sqrt{\Delta m^2_{atm}}$,  Eqn.~\ref{eq:mb} becomes,
\begin{equation}
m_{\beta\beta}=|\sqrt{\Delta m^2_{atm}}c^2_{12}c^2_{13}c^2_{14}+\sqrt{\Delta m^2_{atm}}s^2_{12}c^2_{13}c^2_{14}e^{i\alpha}+\sqrt{2\Delta m^2_{atm}}s^2_{13}c^2_{14}e^{i\beta}+\sqrt{\Delta m^2_{LSND}}s^2_{14}e^{i\gamma}|.
\end{equation}
Using the representative values as earlier, we obtain,
\begin{equation}
m_{\beta\beta}= \sqrt{\Delta m^2_{atm}}~|0.644+0.356 e^{i\alpha}+0.034 e^{i\beta}+22.80 s^2_{14}e^{i\gamma}|.
\end{equation}
So in this case the cancellation occurs since $\sin\theta_{14}$ can take values in the range $0.08-0.17$.
\end{itemize}

\section{Sensitivity in the future experiments}\label{section4}

The future generation $0\nu\beta\beta$ experiments are intending to probe the region $m_{\beta\beta} \sim 10^{-2}$ eV. These experiments include LEGEND, SuperNEMO, CUPID, CUORE, SNO, KamLAND-Zen, nEXO, NEXT, PandaX etc. (See \cite{Agostini:2017jim} for a review). A positive signal in these experiments could be due to IH (three generation or 3+1 generation) or due to NH (3+1 picture) for both LMA and DLMA solutions. If these experiments give a negative result, the next generation of experiments have to be designed with a sensitivity range of $10^{-3}$ eV \cite{Pascoli:2007qh,Penedo:2018kpc}.

\begin{figure}

    \includegraphics[scale=0.63]{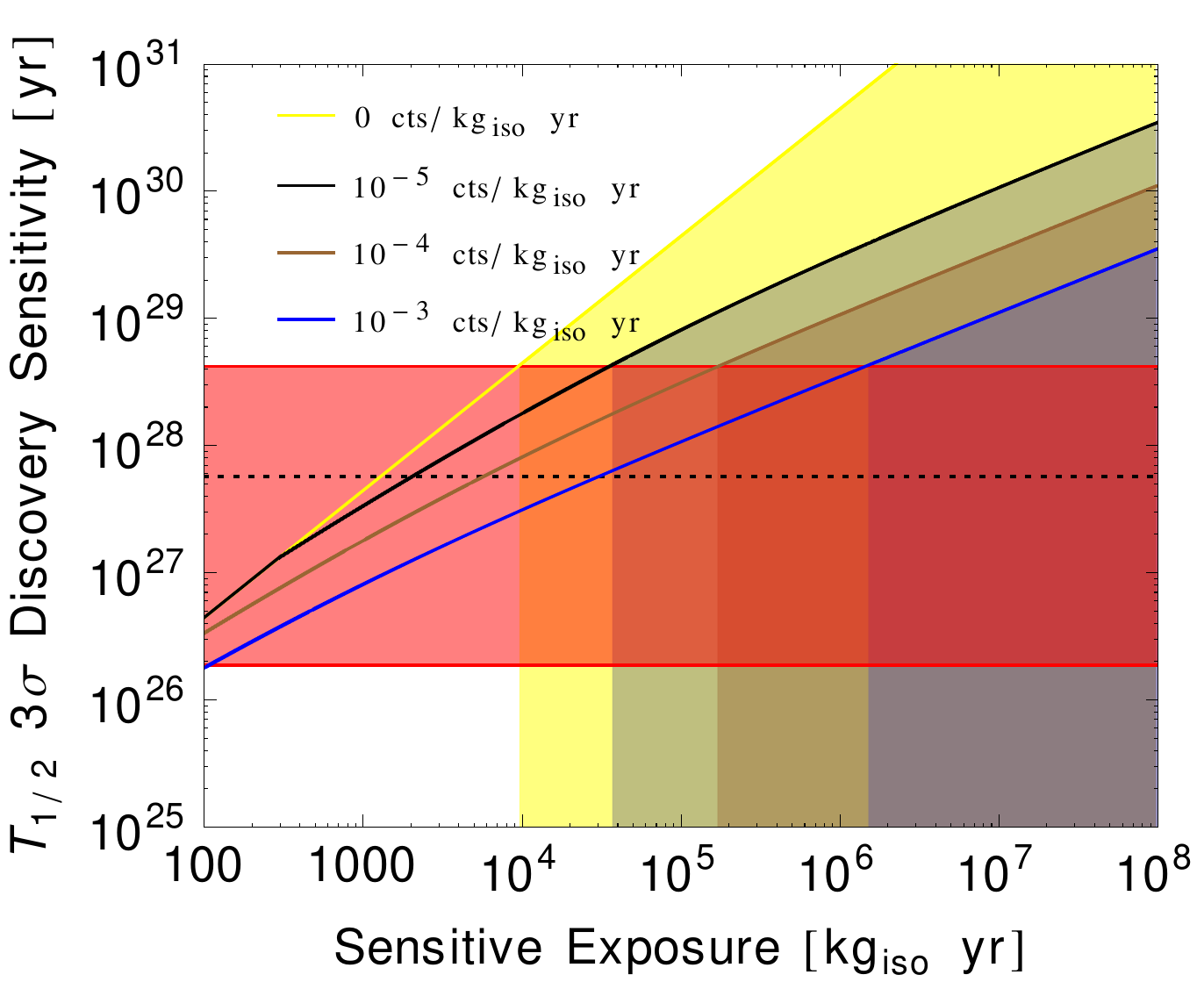}
    \caption{  ${}^{136} Xe$ discovery sensitivity as a function of 
sensitive exposure for different sensitive background levels. The yellow, black, brown and blue lines correspond to four different values of the sensitive background levels as shown in the figure.} \label{Xe}
\end{figure}

In this section, we calculate the sensitivity in the future ${}^{136}Xe$ experiments for which we have adopted the method discussed in reference \cite{Agostini:2017jim}. The value of $T_{1/2}$ for which an experiment has a $50\%$ probability of measuring a $3\sigma$ signal above the background is defined as the $3\sigma$ discovery sensitivity of $T_{1/2}$. It is given as,
\be  T_{1/2} = \textrm{ln}2 \frac{N_A \epsilon}{m_a S_{3\sigma}(B)} .\ee
In this equation, $N_A$ is the Avogadro number, $m_a$ is the atomic mass of the isotope, and $B = \beta\epsilon$ is the expected background where, $\epsilon$ is the sensitive exposure and $\beta$ is the sensitive background. $S_{3\sigma}$ is the value for which half of the measurements would give a signal above $B$ for a Poisson signal and this can be obtained from the equation,
$$1-{CDF}_{Poisson} (C_{3\sigma}|S_{3\sigma}+B) = 50 \%.$$
$C_{3\sigma}$ stands for the number of counts for which the cumulative 
Poisson distribution with mean as $B$ obeys,
$$CDF_{Poisson}(C_{3\sigma}|B) = 3\sigma.$$ 
 We use the  normalized upper incomplete gamma function to define $CDF_{Poisson}$ as a continuous 
distribution in $C$ as follows,
$$ CDF_{Poisson}(C|\mu) = \frac{\Gamma(C+1,\mu)}{\Gamma(C+1)}. $$
This avoids the discrete variations that would arise in the discovery 
sensitivity if $C_{3\sigma}$ is restricted to be integer valued.
Using the above equations, we have calculated the $T_{1/2}$ discovery sensitivities of 
$0\nu\beta\beta$ as a function of $\epsilon$ for various values of 
$\beta$ for  ${}^{136}Xe$ nucleus and the results are shown in Fig.\ref{Xe}.

%%%%%%%%%%%%%%%%%%%%%%%%%%%%%%%%%%%%
%%%%%%%%%%%%%%%%%%%%%%%%%%%%%%%%%%%%

 In this plot, the red shaded band corresponds to the new allowed region of $m_{\beta \beta} \sim 0.008 -0.04$ eV  for the 
DLMA solution for the NH case with a sterile neutrino. This band in $m_{\beta\beta}$ which is due to the variation of the parameters in the PMNS matrix, is converted to a band in $T_{1/2}$ using Eqn.~\ref{0nbbthalf}, by taking into account the NME uncertainty as given in Table \ref{Isotopes}. The dotted black line corresponds to the future $3\sigma$ sensitivity of nEXO, which is $T_{1/2} = 5.7 \times 10^{27}$ years \cite{Kharusi:2018eqi}. The yellow, black, brown and blue lines correspond to four different values of the sensitive background levels of $0$, $10^{-5}$, $10^{-4}$ and $10^{-3}$ $~\textrm{cts}/(\textrm{kg}_{\textrm{iso}} \textrm{yr})$ respectively. From the figure, we can see that a large part of this newly allowed region for NH is in the reach of the nEXO experiment.
With lower background levels and/or higher sensitive exposure, the next generation experiments can probe this entire region.

In Table \ref{Isotopes}, we have given the $T_{1/2}$ ranges corresponding to the newly allowed region of $m_{\beta \beta}$ for the 
DLMA solution for the NH case with a sterile neutrino , i.e., $m_{\beta \beta} = 0.008 - 0.04$ eV for three different isotopes. The predictions for $T_{1/2}$  is highly affected by the uncertainties in NMEs. To calculate the matrix element, one can use different models like shell model, quasiparticle random-phase approximation (QRPA), interacting boson model (IBM), etc., and each model has their own advantages and disadvantages \cite{Engel:2016xgb}. The inverse half life also depends on the fourth power of the weak axial vector $(g_A)$. Hence a small uncertainty in $g_A$ largely affects the extracted value of $m_{\beta\beta}$ from observed value of $T_{1/2}$. It depends on the mass number of the nucleus and the momentum transfer. The quenching of $g_A$ from its free nucleon value arises due to nuclear medium effects and nuclear many body effects. The detailed study of $g_A$ and its possible uncertainties are discussed in \cite{Suhonen:2017krv}. In our work, we have used those values of $M_\nu$ for which $g_A$ = 1.25 \cite{Engel:2016xgb,Agostini:2018tnm}.

 %%%%%%%%%%%%%%%%%%%%%%%%%%%%%%%%%%%%%%%%%%%%%%%%%%%%%%%%%%%%%%%
%%%%%%%%%%%%%%%%%%%%%%%%%%%%%%%%%%%%%%%%%%%%%%%%%%%%%%%%%%%%%%
\begin{table}[ht]
 $$
 \begin{array}{|c|c|c|c|}
 \hline \textrm{Isotope}& \textrm{NME}~(M_\nu) & G (10^{-15} \textrm{year}^{-1}) & T_{1/2} \, \textrm{range (years)}  \\
  
 \hline
 
  {}^{136}Xe  & 1.6-4.8 &  14.58   & 1.87\times 10^{26}- 4.20\times 10^{28} \\
  
  \hline
 
 {}^{76}Ge  & 2.8-6.1 &   2.363   &  7.13 \times 10^{26} - 8.47 \times 10^{28}    \\
   
 \hline
 
  {}^{130}Te  & 1.4-6.4 &    14.22   &  1.08 \times 10^{26} - 5.63 \times 10^{28}    \\
   
 \hline
 
 \end{array}
 $$
\caption{\small{ The $T_{1/2}$ ranges corresponding to the DLMA region $m_{\beta\beta} = 0.008-0.04$ eV, the new allowed region for the 
DLMA solution for the NH case with a sterile neutrino for different isotopes.  The NME values \cite{Engel:2016xgb,Agostini:2018tnm} and the phase space factors \cite{Kotila:2012zza} used in the calculation are also given.
 }}\label{Isotopes}\end{table}

%%%%%%%%%%%%%%%%%%%%%%%%%%%%%%%%%%%%%%%%%%%%%%%%%%%%%%%%%%%%%%%
%%%%%%%%%%%%%%%%%%%%%%%%%%%%%%%%%%%%%%%%%%%%%%%%%%%%%%%%%%%%%%

\section{Summary}\label{section5}

In this paper, we have studied the implications of the DLMA solution to the
solar neutrino problem for $0\nu\beta\beta$ in the 3+1 scenario, 
including an additional sterile neutrino. 
We have verified that even in the presence of sterile neutrino, 
the MSW resonance can take place in the DLMA region. 
Next, we have studied how for these values of $\theta_{12}$, 
the predictions for $0\nu\beta\beta$ in 3+1 picture is changed
as compared to the predictions for 3+1 scenario
assuming ordinary LMA solution. 
We also compare with the predictions of $m_{\beta\beta}$ for the three generation picture.

We find that for IH, there is no change in $m_{\beta\beta}$ predictions as compared to the 3+1 case assuming $\theta_{12}$ to be in the standard LMA region. 
This is because in this case, the maximum value of $m_{\beta\beta}$ is 
independent of $\theta_{12}$ and the minimum value of $m_{\beta\beta}$ 
is of the order of $\sim 10^{-4}$ eV where the difference is not very evident. 
In particular, the cancellation region which was reported earlier for 
3+1 sterile neutrino picture also continues to be present for the 
DLMA parameter space due
to the 
contribution from the fourth mass eigenstate. 
This conclusion is similar to the conclusion obtained for the three generation 
case, for which also the LMA and DLMA solutions gave same predictions for 
$m_{\beta\beta}$ in the case of IH. 

In the case of NH, cancellation can occur for certain values of 
$m_{lightest}$ and the values for which this happens
is higher for the DLMA solution.  
Also, the maximum value of $m_{\beta\beta}$ is 
same for the standard LMA and DLMA solutions in the 3+1 scenario 
and unlike the three generation case there is no desert region between 
NH and IH. 
However, the maximum value  is higher  than that for the  
three generation DLMA case. 

If future experiments  
with sensitivity reach of $\sim$ 0.015 eV observe a positive signal 
for $0\nu\beta\beta$ 
then it could be due to IH (three generation or 3+1 generation) 
 or due to NH (3+1 picture)  for both 
LMA and DLMA solutions. 
If however, no such signal is found  then for three generation picture 
$0\nu\beta\beta$ experiments can disfavor IH  
and one moves to the next frontier
of $0.001$ eV \cite{Pascoli:2007qh,Penedo:2018kpc}. 
In this regime a demarcation between  
LMA and DLMA is possible 
for three generation picture if a signal is obtained for $m_{\beta \beta} 
\gtrsim 0.004$ eV  \cite{N.:2019cot}.        
However, if the sterile neutrino hypothesis is true then distinction between
NH and IH is not possible from $0\nu\beta\beta$ experiments. 
This also spoils the sensitivity to demarcate between LMA and DLMA 
solutions.  If however, the current indication of NH from 
accelerator experiments  is confirmed by future data then 
the next generation of $0\nu\beta\beta$ 
experiments with sensitivity reach up to $10^{-3}$ eV can distinguish 
between LMA and DLMA solution in presence of a sterile neutrino  
for $m_{lightest} \lesssim 0.005$ eV.

%%%%%%%%%%%%%%%%%%%%%%%%%%%%%%%%%%%%%%%%%%%%%%%%%%%%%%%%%%%%%%%%%%%%%%%%%%%%%%%%%%%%%%%
%\bibliographystyle{plain}
%\bibliographystyle{unsrt}
%\bibliographystyle{apsrev4-1}
%\bibliographystyle{apsrev}
%\bibliographystyle{decsci}
\bibliographystyle{utphys}
\bibliography{nlbdk}

\end{document}